\documentclass[twocolumn]{aa}
\usepackage{graphicx}
\usepackage{natbib}

\def\lsim{\raise0.3ex\hbox{$<$}\kern-0.75em{\lower0.65ex\hbox{$\sim$}}} 
\def\gsim{\raise0.3ex\hbox{$>$}\kern-0.75em{\lower0.65ex\hbox{$\sim$}}} 
\def\lesssim{\mathrel{\hbox{\rlap{\hbox{\lower4pt\hbox{$\sim$}}}\hbox{$<$}}}}
\def\gtrsim{\mathrel{\hbox{\rlap{\hbox{\lower4pt\hbox{$\sim$}}}\hbox{$>$}}}}
\def\sc1{\raise2.1ex\hbox{\tiny $r\!\!=\!\!4$}\kern-0.95em{\hbox{$=$}}}

\newcommand{\mamo}[1]{\mbox{$#1$}}
\newcommand{\unit}[1]{\ifmmode \:\mbox{\rm #1}\else \mbox{#1}\fi}

\newcommand{\mone}{\mamo{^{-1}}}            
\newcommand{\km}{\unit{km}}
\newcommand{\kms}{\unit{km~s\mone}}

\def\ltsima{$\; \buildrel < \over \sim \;$}
\def\simlt{\lower.5ex\hbox{\ltsima}}
\def\gtsima{$\; \buildrel > \over \sim \;$}
\def\simgt{\lower.5ex\hbox{\gtsima}}          

\def\sc{{\rm Science\ }}

\begin{document}

\title{Geometrical   tests of     cosmological  models}
\subtitle{III. The cosmology-evolution diagram at $z=1$}

\author{
C. Marinoni$^{1}$, A. Saintonge$^{2}$, T. Contini$^{3}$, C.J. Walcher$^4$,
R. Giovanelli$^{2}$, M.P.  Haynes$^{2}$, K.L. Masters$^{5}$,\\
O. Ilbert$^{4}$, A. Iovino$^6$, V. Le Brun$^{4}$, O. Le Fevre$^{4}$,
A. Mazure$^{4}$, L. Tresse$^{4}$, J.-M. Virey$^{1}$, 
S. Bardelli$^7$, D. Bottini$^8$, B. Garilli$^8$, G. Guzzo$^9$, D. Maccagni$^8$, 
J.P. Picat$^3$, R. Scaramella$^9$, M. Scodeggio$^8$, P. Taxil$^{1}$, 
G. Vettolani$^{10}$, A. Zanichelli$^{10}$, E. Zucca$^{7}$ }

\offprints{C. Marinoni, \email{marinoni@cpt.univ-mrs.fr}}

   \institute{
$^1$ Centre    de   Physique  Th\'eorique\thanks{Centre  de   Physique
Th\'eorique is UMR 6207 -
 ``Unit\'e  Mixte de Recherche'' of  CNRS and of the Universities ``de
 Provence'', ``de  la M\'editerran\'ee''  and  ``du Sud  Toulon-Var''-
 Laboratory affiliated to  FRUMAM  (FR  2291).}, CNRS-Universit\'e  de
 Provence, Case 907, F-13288 Marseille, France.\\
$^2$ Department of  Astronomy,  Cornell University, Ithaca,  NY 14853,
USA\\    $^3$  Laboratoire        d'Astrophysique   de  l'Observatoire
Midi-Pyr\'en\'ees,   UMR  5572,     31400 Toulouse,   France   \\ $^4$
Laboratoire d'Astrophysique  de Marseille, UMR 6110, CNRS Universit\'e
de   Provence, 13376   Marseille, France  \\  $^5$ Harvard-Smithsonian
Center   for    Astrophysics,   Cambridge,   MA   02143,   USA\\  $^6$
INAF-Osservatorio Astronomico di Brera - Via Brera 28, I-20121 Milano, Italy\\
$^7$ INAF-Osservatorio    Astronomico  di Bologna   -  Via  Ranzani 1,
Bologna,  Italy\\ $^8$  IASF-INAF -  via  Bassini 15, I-20133, Milano,
Italy\\ $9$ INAF-Osservatorio Astronomico di Brera - Via Bianchi 46, 
I-23807 Merate, Italy\\
$^{10}$ IRA-INAF - Via Gobetti,101, I-40129, Bologna, Italy\\ }

\authorrunning{Marinoni et al.}  \titlerunning{Geometrical tests of cosmological models.III.}

\date{Received 17 January 2007 / Accepted 6 september 2007}

\abstract{
The  rotational velocity  of  distant galaxies,  when interpreted as a
size  (luminosity) indicator, may be   used as a   tool to select high
redshift standard rods (candles)  and probe world models and  galaxy
evolution  via the  classical   angular diameter-redshift  or  Hubble diagram
tests. We implement the proposed testing strategy 
using a sample of $30$ rotators spanning the redshift range $0.2<z<1$
with high resolution spectra and images obtained  
by the VIMOS/VLT Deep Redshift Survey (VVDS) and the Great Observatories 
Origins Deep Survey (GOODs). We  show that by applying at the same 
time the angular diameter-redshift and  Hubble diagrams  to the same
sample of objects (i.e. velocity selected galactic discs) one can  derive a 
characteristic chart, the cosmology-evolution diagram, 
mapping the relation between   global cosmological  parameters  and local   
structural  parameters  of discs  such  as  size  and luminosity.
This chart allows to put constraints on cosmological parameters when  general 
prior information about discs evolution is available. In particular, by 
assuming that equally rotating large discs cannot be less luminous at $z=1$ than at present 
($M(z=1)  \lesssim M(0)$),  we find that a
flat matter   dominated  cosmology ($\Omega_m=1$)   is  excluded  at a
confidence level of $2 \sigma$  and an open   cosmology with low  mass
density ($\Omega_m \sim    0.3$)  and  no dark energy     contribution
($\Omega_{\Lambda}$) is excluded at a confidence level greater than $1
\sigma$. Inversely, by assuming prior knowledge about the cosmological model, 
the cosmology-evolution diagram can be used to
gain useful insights about the redshift evolution of  
baryonic discs hosted in dark matter halos of nearly {\it equal  masses}.
In particular, in a $\Lambda CDM$ cosmology,  we find
evidence  for  a bimodal   evolution   where the low-mass  discs  have
undergone significant surface brightness  evolution over the  last 8.5
Gyr, while more massive systems have not. We suggest that this dichotomy
can    be  explained  by  the epochs   at  which   these two different
populations last assembled.


}

\maketitle

\section{Introduction}

Deep redshift surveys of  the Universe, such as    the
VIMOS/VLT    deep  redshift  survey     (VVDS, \citet{lef05}) and  the
ACS/zCOSMOS survey  \citep{lil06} are currently  underway to study the
physical properties of high   redshift  galaxies. Motivated by these
major   observational efforts, we are currently exploring whether  
high redshift galaxies can also be used as cosmological tracers.
Specifically, we are trying to figure out if these
new and large sets of spectroscopic data can be meaningfully used
to probe, in a geometric way, the value of the constitutive parameters of
the Friedmann-Robertson-Walker cosmological model.

A whole arsenal of classical geometrical methods has been developed to
measure global properties of the universe.  The central feature of all
these tests is the attempt  to probe the various operative definitions
of relativistic distances by means of scaling relationships in which a
distant dependent
 observable, (e.g. an angle or a flux),
is expressed  as a function of  a distance independent fixed
quantity (e.g. metric size or absolute luminosity).

A  common  thread of  weakness   in all  these  approaches to  measure
cosmological parameters using distant  galaxies  or AGNs selected   in
deep  redshift surveys is  that there are no   clear criteria by which
such cosmological objects should be considered universal standard rods
or standard candles.

Motivated   by  this,   in    previous  papers   (\citet{mar04}    and
\citet{mar07}, hereafter Paper I) we have investigated the possibility
of  using the  observationally   measured and theoretically  justified
correlation between size/luminosity  and disc  rotation velocity as  a
viable  method  to  select  a set   of  high  redshift galaxies,  with
statistically  homologous dimensions/luminosities. This set of tracers
may be used to  test the evolution of the  cosmological metric via the
implementation  of the standard   angular diameter-redshift  
and  Hubble diagram tests.

Finding valid standard rods, however, does not solve the
whole   problem; the  implementation of  the angular
diameter-redshift test using distant galaxies  is hampered  by the difficulty  of
disentangling the  effects of galaxy  evolution  from the signature of
geometric expansion of the universe.

In Paper  I we have  determined  some general conditions  under  which
galaxy kinematics    may be   used  to  test  the  evolution    of the
cosmological  metric.  We  have shown that   in the particular case in
which disc evolution  is linear and modest  ($<$30$\%$ at $z=1.5$), the
inferred values of the dark energy density parameter $\Omega_Q$ and of
the cosmic equation  of  state  parameter  $w$ are  minimally   biased
($\delta  \Omega_Q=\pm    0.15$  for  any $\Omega_Q$     in  the range
$0<\Omega_Q<1$).

In Paper I, we also looked  for  cosmological  predictions   that  rely on less
stringent assumptions, i.e.  which  do not require  specific knowledge
about  the particular functional  form    of the standard   rod/candle
evolution.   In  particular,   we  showed   how  
velocity-selected rotators  may be used to construct   a
{\it cosmology-evolution  diagram} for disc galaxies. This is a chart mapping the local  
physical parameter space of rotators (absolute luminosity and disc linear size) onto the 
space of  global, cosmological parameters  ($\Omega_{m}$, $\Omega_{Q}$).
Using this diagram it  is possible  to  extract
information about cosmological parameters once the amount of size/luminosity 
evolution at some reference epoch is known.
{\it Vice-versa}, once  a cosmological model is assumed, the cosmology-evolution
mapping may  be  used  to  directly infer   the  specific  time evolution  in
magnitude and size of disc galaxies that are hosted {\it in dark matter halos
of similar mass}. 

We stress that this last way of reading the cosmology-evolution diagram 
offers a way to explore galaxy evolution  which is orthogonal to more traditional  methods.
In particular, insights  into  the mechanisms of   galaxy evolution are traditionally
accessible through the study of disc galaxy scaling relations, such as
the   investigation   of    the   time-dependent  change      in   the
magnitude-velocity    (Tully-Fisher)          relation          (e.g.,
\citet{vog96,bohm04,bam06}), of  the  magnitude-size relations  (e.g.,
\citet{lil98,sim99,bou02,bar05}),  or    of    the   disc  ``thickness"
\citep{res03,elm05}.   By applying the angular size-redshift test
and the Hubble diagram to velocity-selected rotators,  we aim at tracing the evolution 
in linear size, absolute magnitude and intrinsic surface brightness 
of disc galaxies that are hosted halos of the same given mass at every cosmic epoch
explored.

In this paper, we present a pilot observational program 
that allowed us to test whether galaxy rotational velocity can be used 
to select standard rods,  
and to derive the cosmology-evolution diagram for disc galaxies 
at redshift $z=1$. 
Our observing strategy was to  follow-up in medium resolution spectroscopic mode
with VIMOS a  set  of emission-line objects selected from a sample of 
galaxies in the Chandra Deep Field South (CDFS) region for which  
high resolution photometric parameters were available
\citep{gia04}.

The outline  of the paper  is as follow: in  \S 2 we describe the VVDS
spectroscopic    data taken in  the CDFS 
region. In \S 3 we outline  a strategy to  test the consistency of the
standard   rod/candle   selection.     In  \S  4      we  derive   the
cosmology-evolution diagram for our sample of rotators, and in \S 5 we
present our  results about disc size,  luminosity  and surface brightness
evolutions.  Discussions and conclusions are presented in
\S  6  and \S 7, respectively.  Throughout, the  Hubble constant  is
parameterized via $h_{70} = H_0/(70 km s^-1 Mpc^-1).$ 
All  magnitudes in this paper are
in the AB  system  \citep{oke83}, and from now the AB suffix will be omitted.

\section{Sample: observations and data reduction}

Our strategy to obtain  kinematic information for the largest possible
sample  of rotators  at   high redshift  was  to re-target  in  medium
resolution  mode (R=2500) galaxies   in the CDFS  region  for  which a
previous pass  in low-resolution  mode \citep{lef04} already  provided
spectral information  such as  redshifts,  emission-line   types,
and equivalent  widths, for galaxies down to I=24. Galaxies were  selected  
as  rotators  if their  spectra  was blue and
characterized  by   emission   line features    (OII, H$\beta$,  OIII,
H$\alpha$).  CDFS photometry was  then  used to confirm the  disc-like
nature of their light distribution (i.e. the absence of any  strong
bulge component), and also to avoid including in the sample  objects
with peculiar morphology or undergoing merging or interaction events.

The final sample of candidates for medium resolution re-targeting was
defined   by  further requiring  that the inclination of the galaxy 
be greater than 60$^{\circ}$ to minimize
biases  in velocity estimation, and that its  identified emission line fall on 
the  CCD under the  tighter constraints imposed by  the
medium resolution grism.   Once   the  telescope pointing  and    slit
positioning were  optimized using the low-resolution spectral information,
the remaining space  on the focal plane  mask was blindly assigned  to
galaxies in the field.

Spectroscopic observations have been obtained with the
    VIMOS spectrograph on the VLT Melipal telescope in October
    2002. The slit width was 1 arcsecond giving a spectral
    resolution R=2500 as measured on the FWHM  of arc lines. Using the
    VIMOS  mask design software and  capabilities  of the slit-cutting
    laser machine \citep{bot05}, slits have been placed on each galaxy
at a position angle aligned with the major axis. The seeing at the time
    of observations was
    0.8 arcseconds FWHM with an integration time of 1.30h 
    split in three exposures of 30 minutes each.
 
Most of the galaxies  in the CDFS area  surveyed by the VVDS have high
resolution images taken   with the ACS  camera  of the  HST  by GOODs.
Images are  available in four  different filters (F435W, F606W, F775W,
F850LP) noted hereafter B,V,I and Z, respectively. A  small fraction of the
targeted galaxies has only I band images  provided by the ESO Imaging
Survey \citep{arn01}.

The   galaxy   rotational  velocity  has   been   estimated  using the
linewidths of the emission lines.     A detailed  analysis  of  the   velocity
extraction algorithm and  of the potential  systematic errors implicit
in this technique  are presented in  Paper II of this serie \citep{san07}. 
This technique to measure rotation velocities imposed itself since many 
galaxies  at high redshift  were too small to measure rotation 
curves reliably, and since summing all the light to form velocity histograms 
increased  the signal-to-noise ratio (S/N) of the detected lines.

Magnitudes have been  computed in the I band and a  K-correction was
applied (see \citet{ilb05} for a  detailed discussion). They were also
corrected  for galactic absorption using  the maps of \citet{sch98} in
the CDFS region  (i.e. on average  a correction of $\sim 0.0016$), and
for galaxy inclination by adopting a standard empirical description of
internal  extinction   $A_{\lambda}$   in  the  pass-band   $\lambda$,
$\gamma\log(\sec i)$, where $i$ is the  galaxy inclination angle as
calculated from the   galaxy axis ratio  and  $\gamma_I=0.92+1.63(\log
2v-2.5)$ \citep{tul98} where $v$ is the maximum rotational velocity of 
a galaxy.

Galaxy sizes have  been specified in  terms of the half-light diameter
(HLD) inferred  in the I band.  Typical errors in the measurements are
$\sigma_{\theta}\sim 0.04"$.  In Paper I we stressed the importance of
using a metric  rather  than an isophotal definition of  galaxy diameters
for cosmological purposes  \citep[e.g.][]{san95}.   We
also verified that the HLDs for our sample of galaxies do not depend on 
wavelength; there is no systematic difference  in the inferred metric diameters
when the HLD is  computed in the  B, V, I or z
filters (see also \citet{san90}, \citet{deJon96}).  
The scatter in the HLDs inferred in different bands is of order 0.02" and 
therefore small in comparison to
the observational uncertainties $\sigma_{\theta}$.

[OII] linewidths have  been translated into  an estimate of the galaxy
rotational velocity, $v$, as detailed in \S3.2 of Paper II. Rotational
velocity   was   derived using  [OII](3727\AA)  lines    (24 objects),
[OIII](5007\AA) lines    (10    objects) and   H$\alpha$(6563\AA)   (5
objects).  23 galaxies have velocities in  the range $0<v(\kms) \leq 100$ (with
mean  velocity of the   sample $\sim 60$ \kms)   and 16 galaxies  have
velocities in the range  $100<v(\kms) \leq  200$ (with  mean velocity  of the
sample $\sim 143$ \kms) .

After data reduction, we were left with a sample of  39 objects, 27 of
which have high resolution imaging. As for the remaining objects with ground
photometry, we only consider in the  following those with $z<0.2$, in order to 
exclude faint and small galaxies for which the size measurements are severely 
compromised by seeing distortions. Therefore, our final "science" sample contains 30 objects.

Data  are organized  and presented  in  Table 1 as  follows: $col. 1:$
galaxy ID in  the EIS catalog, $col. 2:$  redshift,  $col. 3$ rotation
velocity,  $col. 4:$  half-light angular radius,  $col. 5:$ magnitude,
$col. 6:$ surface brightness within the half-light radius.

\begin{center} \begin{table*} \begin{tabular}{c c c c  c c} \hline EIS
ID &  $z$ &  $v$(\kms)  & $\theta^{o}(arcsec)$  & $m^{o}$ &  $\mu^{o}$
(mag/arcsec$^2$) \\ \hline  30445 & 0.9332  & 97 &   0.180 & 23.744  &
22.03\\ 32177 & 0.8934 & 68 & 0.149 & 23.681  & 21.54\\ 31328 & 0.4164
& 79 & 0.606 & 22.297 & 23.20\\ 32998 & 0.1464 & 28 & 0.857 & 20.794 &
22.45\\ 34826 & 0.4559 & 204 & 0.715 & 22.684 & 23.95\\ 34244 & 0.5321
& 55 & 0.298 & 22.324 & 21.69\\ 29895 & 0.6807 & 44 & 0.220 & 23.571 &
22.23\\ 37157 & 0.8677 & 129 & 0.751 & 23.399 & 24.77\\ 33200 & 0.1267
& 96 & 0.806 & 20.519 & 22.04\\ 33763 & 1.0220 & 130  & 0.755 & 23.404
&  24.79\\ 31501 &  1.0360 &  140 & 0.747  &  23.513 & 24.87\\ 31194 &
0.3320 & 80 & 0.818 & 21.950 & 23.51\\  29342 & 0.4680 &  55 & 0.481 &
23.565 & 23.97\\ 29232 & 0.8610 & 155 & 0.370 & 22.320 & 22.15\\ 34325
& 0.3334 & 26 & 0.221 & 23.570 & 22.20\\ 34560 & 0.8618 & 28 & 0.204 &
23.549 & 22.00\\ 36484 & 0.7539 & 25 & 0.202 &  23.781 & 22.30\\ 16401
& 1.1000 & 306 & 0.913 & 21.628 & 23.42\\ 17811 & 0.8143 & 170 & 0.726
&  23.573 & 24.87\\ 17362 &  0.6814 & 178 &   0.683 & 22.379 & 23.54\\
22685 & 0.8411 & 115 & 0.891 & 22.147 & 23.88\\ 17255  & 0.1787 & 99 &
1.203 &  21.758 & 24.16\\   15152 & 0.7931 &  169  & 0.895 &  21.702 &
23.46\\ 15099 & 0.3661 & 70 & 0.509 & 21.171 &  21.70\\ 19702 & 0.6770
& 62 & 0.295 & 22.812 & 22.16\\ 16377 & 0.5621 & 36 & 0.490 & 22.944 &
23.37\\ 17421 & 0.7834 & 99 & 0.361 &  23.280 & 23.06\\ 20202 & 0.5763
& 26 & 0.190 & 23.318 & 21.70\\ 18416 & 0.8859 & 99 & 0.205 & 23.637 &
22.18\\ 17534 & 0.3493 & 35 & 0.271 & 23.837  & 22.99\\ 15486 & 0.6613
& 146 & 0.551 & 22.495 & 23.20\\ 19684 & 0.8588 & 104 & 0.424 & 22.790
& 22.92\\ 18743  & 0.6800  & 81  & 0.447  & 22.950  & 22.70\\ 15553  &
0.4584 & 183 & 0.991 & 19.292 & 21.26\\ 18417 & 0.5350  & 36 & 0.342 &
22.877 & 22.54\\ 18779 & 0.5623 & 59 & 0.351 &  22.445 & 22.15\\ 21252
& 0.5795 & 102 & 0.414 & 22.854 & 22.93\\ 20708 & 0.1228 & 166 & 1.540
& 18.427 &  21.40\\ 18853 &  0.6509 & 116  & 0.556 &  21.350 & 22.10\\
\hline  \end{tabular}   \caption{Properties    of  the  Galaxy Sample}
\label{table1} \end{table*} \end{center}

\section{Selection of Standard rods/candles}

An observable relationship exists  between the metric radial dimension
$D$  of a disc  and its speed  of rotation $v$. An analogous empirical
relationship  connects rotation  with  luminosity  \citep{tul77}.   In
Paper  I we  have proposed to  use   information on the  kinematics of
galaxies, as encoded  in  their   OII  emission-line width,  to   {\it
objectively} identify  standard rods/candles  at   high redshifts.   A
discussion of the   requirements  and of  the  optimal  strategies  to
fulfill this observational program is detailed in Paper I.

A variety of  standard rod candidates have  been  explored in previous
attempts of  providing a direct  geometrical proof of the curvature of
the universe.   A  common thread of weakness  in  all these attempts is
that there are no clear physical nor statistical criteria by which the
proposed objects (clusters, extended radio lobes or compact radio jets
associated with  quasars  and AGNs)   should be considered   universal
standard rods/candle.

Even assuming that a particular class of standards is identified, the
length  of the rod  remains unknown.  Since  the inferred cosmological
parameters heavily  depend on the  assumed value  for the object  size
\citep{lim02}, an {\it  a-priori} independent statistical study of the
standard rod absolute calibration is an imperative prerequisite. 
In Paper II, we used a large sample of galaxies from the 
SFI++ catalog  \citep{spr07}  to fix the local calibration values  for 
absolute magnitudes  and  linear diameters  of galaxies with a given 
rotational velocity.

\subsection{Velocity selection of rotators: test of consistency}

We have  seen that, in order to implement the proposed test, we need two sample of rotators: 
the ``data sample" (galaxies  with the same  rotational velocity 
selected over  the widest possible redshift range; the sample presented in \S 2), 
and the ``calibration sample" (rotators at redshift $z\sim 0$ for which the physical size of the 
linear diameter is known; the SFI++ sample analyzed in Paper II). 
This last sample allows us to calibrate the zero-point of the Hubble and angular
size-redshit diagrams (i.e. $M_v(0)$ and $D_v(0)$ in  eqs. 2 an 4).

We stress that the disc rotational velocity of galaxies 
in the two samples is measured using two different velocity 
indicators (spectroscopic lines) and two different velocity extraction 
methods. Specifically we use  OII linewidths to measure the rotational 
velocity of the distant ``data" sample and $H_{\alpha}$ rotation curves 
to measure the velocities of the local ``calibration" galaxies. 
Therefore, it is imperative to check  that  possible   biases or errors
introduced by  combining   velocities  inferred  using  systematically
different measuring  techniques do   not prevent a
meaningful comparison between different samples at different redshifts.

To this purpose  we  have implemented the following  testing strategy.
Given a spectroscopically-selected sample of standard candles $M_v(0)$
with  rotational velocity $v$,  one  can derive the observed  apparent
magnitude $m^{o}$  of a  standard candle  located  at redshift $z$, by
using      the   relation        \citep{SAN}:   
\begin{equation}
m^{o}=m^{th}(M_v(0),z,\vec{p})+\epsilon_M(z,\vec{p})+K(z)  \label{eqa}
\end{equation}    
where       
\begin{equation}
m^{th}=M_v(0)+5\log d_{L}(z,\vec{p})+25      
\label{eqb}    
\end{equation}     
and  where $d_{L}(z,\vec{p})$ is the luminosity distance (depending on the set of
cosmological parameters  $\vec{p}$),  $K(z)$ is the K-correction  term
and $\epsilon_M(z,  \vec{p})$  is the  {\it  a-priori} unknown cosmology-dependent 
evolution      in    luminosity     of     our      standard   candle,
i.e.  $\epsilon_M(z,\vec{p})=M_v(z,\vec{p})-M_v(0)$ is the  difference
between the  absolute magnitude of an object  of rotational velocity $v$
measured at redshift $z$ with respect to the un-evolved local standard
value  $M_v(0)$.

Similarly, one can parameterize  any possible evolution 
affecting the standard rod $D_{v}(0)$ by writing its observed apparent subtended angle
at redshift $z$ as
\begin{equation}
\theta^{0}=\theta^{th}(D_v(0),z,\vec{p})[1+\delta(z,\vec{p})]
\label{eqc} 
\end{equation}  
where the  theoretically expected  angular
scaling     ($\theta^{th}$)     is     given    by    
\begin{equation}
\theta^{th}=\frac{D_v(0)}{d_{A}(z,\vec{p})},   \;\;\;\;\;\;  d_{A}=d_L(1+z)^{-2}           
\label{eqd}     
\end{equation}   
and where $\delta(z,\vec{p})$ is a cosmology-dependent function which 
describes the relative redshift  evolution  of the standard  rod,
i.e $\delta \equiv (D_v(z,\vec{p})-D_v(0))/D_v(0)
\equiv \epsilon_{D}/D_{v}(0)$. We note that any possible  evolution 
in the standard rod angular size is related to the evolution in 
its linear dimension as follows: $\epsilon_{\theta}=\epsilon_D/d_{A}$.
Here and in the following, we assume that the angular size
of fixed-velocity rotators is estimated using the galaxy half-light 
diameter $D_{v}$.

From the definition of  wavelength-specific surface brightness, $\mu$,
we deduce that the variation as a  function of redshift in the average
intrinsic surface brightness  within a radius R for   our  
set of velocity selected galaxies  (i.e. 
$\Delta \langle       \mu^{in}(z) \rangle_{R} \equiv 
\langle       \mu^{in}(z)-\mu^{in}(0) \rangle_{R}$) is 
\begin{equation}
\Delta\langle \mu^{in}(z)\rangle_R=\Delta M_v(<R) +  5 \log \frac{R(z)}{R(0)} 
\end{equation}

By choosing the half-light  diameter $D_v$ as a metric
definition for the size of a standard rod, we immediately obtain the intrinsic 
surface brightness evolution within $D_{v}$ as
\begin{equation}   \Delta  \langle       \mu^{in}(z) \rangle_{D_v}   =
\epsilon_M(z, \vec{p}) + 5 \log (1+\delta(z,\vec{p}))
\label{eqe}. \end{equation}
While the specific amount of evolution in luminosity and 
size do in principle depend on the specific background cosmological model adopted, 
the corresponding evolution in intrinsic surface brightness is a
cosmology-independent quantity.

The evolution in intrinsic  surface brightness is not a directly measurable
quantity, but, in  a FRW metric,  this  quantity can be easily related 
to  the apparent surface brightness change   observed   in a  waveband   
$\Delta \lambda$  by the relation
\begin{equation}   \Delta \langle\mu^{o}(z)\rangle_{D_{v}}=\Delta
\langle  \mu^{in}(z)\rangle_{D_{v}} +2.5  \log(1+z)^4+K(z)      \label{eqf}
\end{equation} We note  that  the left-hand side of  Eq. \ref{eqf} is 
directly measurable using photometric images. Moreover, it can be measured
without  assuming any specific  galaxy light profile  and it will  be, in
general,  a non   linear     function  of  redshift.    By   combining
eqs.   \ref{eqa},  \ref{eqc}, \ref{eqe}   and \ref{eqf}  we define the
$\eta$ function:
\begin{equation}
\eta=m^o(z)-\Delta \langle\mu^{o}(z)\rangle_{D_v}+5\log \theta^o(z).
\label{uno}
\end{equation}

The  specific   combination  in Eq. \ref{uno}   of observed   magnitudes,  half-light
diameters,  and evolution in the observed surface 
brightness  within the HLD ($\Delta
\langle\mu^{o}(z)\rangle_{D_{v}}=\langle\mu^{o}(z)\rangle_{D_v}-
\langle\mu^{o}(0)\rangle_{D_v}$) is,      by     construction,    a
redshift-invariant quantity which is equal to 
\begin{equation}     
\eta=M_v(0)+5\log D_v(0)+25.    
\label{due}
\end{equation}

From  a theoretical point of view, we emphasize that  the  
$\eta$-estimator given in Eq. \ref{uno} does not explicitly depend on i)
K correction, ii)  evolution  in luminosity  or size  of  our standard
sources and iii) on the specific gravitational model assumed to derive
the exact functional form of the angular and luminosity distances.

From an observational point of view, we stress that Eq. \ref{uno}
can be directly estimated using  photometric images of the ``data" sample,
while Eq. \ref{due} may be expressed in terms of the locally measured  
absolute magnitudes and linear diameters of  our  ``calibration" sample. 
Therefore, by simply comparing the values of the $\eta$ function   
inferred using the ``data" sample (Eq. \ref{uno}) with the constant value 
predicted  using the ``calibration" sample (Eq. \ref{due}), we
can test for the presence of eventual  biases in our data. 
The goal is to reveal possible systematics that could be introduced,
for  example,   by  the  different   techniques  with  which  rotation
properties  are inferred locally   (mainly using $H_{\alpha}$ rotation
curves)  and at higher redshift (mainly  using OII line-widths).
Clearly, a mismatch would indicate that our spectroscopic selection 
technique fails  in selecting 
homologous classes of objects embedded in halos of nearly the same mass 
at different redshifts.
 
Since our total sample is still limited, at present it is practical to
implement the proposed test of consistency by defining   only two broad classes of 
velocity-selected galaxies: a
low-velocity  sample of  standard    rods/candles with  $0<v\leq  100$\kms \ 
containing 22 galaxies with mean rotational velocity of $\sim 60$ \kms
\ ($S_{60}$   sample)  and  a  high-velocity    set of  objects   with
$100<v<200$\kms \  containing  8 rotators with   mean velocity of $\sim 143$
\kms \ ($S_{143}$  sample).  The size  (HLD),  absolute luminosity and
mean surface brightness $\mu(0)$ within the  HLD of local galaxies are
derived using the calibration relationships of Paper II and are quoted
in   Table \ref{table2}.   Clearly, with   more  high resolution  data
becoming available,  it will be possible  to split the sample in finer
velocity  bins and thus   select standard rods/candles having  smaller
size/luminosity dispersions.

\begin{table*}     \begin{tabular}{c  c  c    c}   \hline  $\langle  v
\rangle/$(\kms)   &   $D(0)/(h_{70}^{-1}$  kpc)  &   $M(0)-5 \log h_{70}$  &   $\mu(0)$
(mag/arcsec$^2$) \\  \hline $60 $&$  4.30 \pm 2.8  $&$ -18.80 \pm 0.75
$&$ 21.50 \pm 0.74$ \\ $143  $&$ 9.00 \pm  2.5 $&$ -21.40 \pm 0.44 $&$
20.45 \pm 0.70$ \\ \hline \end{tabular} \caption{Local calibration for
diameters, absolute magnitudes  and surface brightness within the half
light radius as derived in Paper II} \label{table2} \end{table*}

\begin{figure} 
\includegraphics[width=90mm,angle=0]{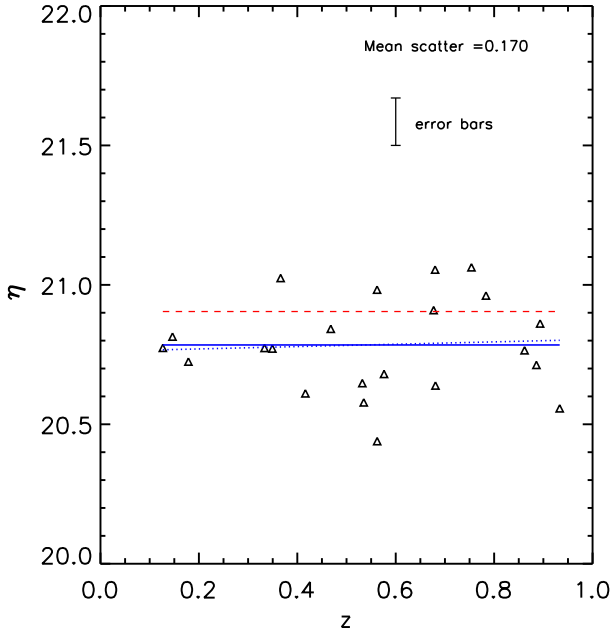}
\caption{The $\eta$ estimator scatter plot  computed using Eq. \ref{uno} 
and  the  $S_{60}$ sample.  The dotted line is the  best fitted linear regression
to the data, while the solid line  represents the best fitted constant
model.  The dashed line represents the estimated values for the $\eta$
function obtained by using locally  calibrated values for the standard
rods and candles,  i.e. using Eq.  \ref{due} with the values specified
in  Table   \ref{table2}. The  best    fitting  linear  regression  is
consistent with being a constant function of redshift.  } \label{fig1}
\end{figure}

In Figure \ref{fig1} we plot the $\eta$-estimator (see Eq. \ref{uno})
for the $S_{60}$ sample of  rotators. The first and third terms on the RHS of Eq.
8 were estimated as explained in \S2, while the second term 
was evaluated  by fitting the observed SB
with a linear model and subtracting from the observation the zero point 
of the model (i.e. the value $\langle \mu^0(0) \rangle$ inferred using the linear model).

A series of conclusions 
can be immediately drawn. First, the  best  fitting  linear regression 
is very well approximated by a constant function of redshift.
This {\it shape} is not only theoretically expected, but it
is consistent with the hypothesis that none of 
the relevant photometric parameters (angular sizes, magnitudes, surface brightnesses) 
measured for our sample of rotators suffer from any redshift-dependent 
systematics.

 Secondly, the consistency of the measurements can be assessed by comparing the 
 {\it scatters}  in $\eta$ estimated by using Eq. \ref{uno} and Eq. \ref{due}. 
The average scatter in Eq. \ref{uno}
measures the  quality with which angular diameters, magnitudes and surface brightnesses have
been measured in  the ``data" sample. This is an extremely useful indicator 
since measuring structural parameters for distant, faint and small galaxies
is not an error-free task. $\sigma_{\eta}$  is thus  a quality
parameter which describes  the overall consistency of our measurements
of the three observables $m^{o}, \mu^{o}$ and $D^{o}$. 
The average scatter in Eq. \ref{due} indicates the robustness with which  local velocity data 
can be used to select standard candles/rods. In other terms
it reflects the intrinsic scatter in the calibration of the Tully-Fisher relation
for local diameters and magnitudes. Clearly, if scatter in Eq. \ref{uno}  
is comparable or bigger than 
scatter in eq \ref{due}, then  our high redshift data would be of low quality 
and definitely useless. The scatter in Eq. \ref{uno}
($\sigma_{\eta}=0.035$) is nearly one order of magnitude smaller than 
that inferred using Eq. \ref{due} ($\sim 0.3$), and, together 
with the absence of any trend in the distribution of the residuals, shows
that the  photometric parameters of the ``data" sample  have been consistently determined
over all the redshift baseline.

Finally,  the {\it normalization} of this constant function tells us 
about the effectiveness of our kinematic measurements (i.e. about the 
homogeneity of the sample of velocity-selected rotators).
The fact that  the $\eta$-value inferred using the ``data" sample  
of rotators with $v=60 \km/s$ (Eq. \ref{uno})  
is well within  the  errors of the $\eta$ value estimated 
using Eq. \ref{due} and our local ``calibration" sample  ($\sim 0.3$)
allows  us to conclude  that both the  high redshift  sample and  the local one  are
homogeneously selected in velocity  space.  
The low redshift counterparts of our rotators have 
a mean luminosity and a mean diameter which combines in Eq. \ref{due}
to give the value which was independently inferred using available local data: 
the high redshift galaxies in  the $S_{60}$ sample are compatible 
with  the hypothesis of being the progenitors of  local galaxies  having  
a standard physical  size of  $D_{60}=4.30  h_{70}^{-1}$ kpc and an absolute
luminosity $M_{60}=-18.80+5 \log h_{70}$, as derived using the SFI++
sample in Paper II.

The consistency test performed using the $\eta$ indicator 
assures us that velocities measured using different methods both
locally and at  high redshift are free of systematics.  Galaxies
with velocity  $v$   at high  redshift   may  actually  have intrinsic
luminosities and diameters   different from those determined for   the
local sample  of   galaxies   with similar  velocity.  But there is no 
evidence against the hypothesis that they are embedded in dark matter 
halos of similar masses. Moreover, if the halo mass does not change across
cosmic time (for example by merging or accretion phenomena),  
galaxies with velocity $v$
estimated using OII linewidths at high redshift will eventually evolve  
into local galaxies having linear size and absolute luminosity 
compatibles with the values predicted by the 
Tully-Fisher relations ($D(v)$ and $M(v)$) locally calibrated using $H_{\alpha}$
rotation curves.

\subsection{Velocity selection of rotators: proof of concept}

\begin{figure*}         
\includegraphics[width=90mm,angle=-0]{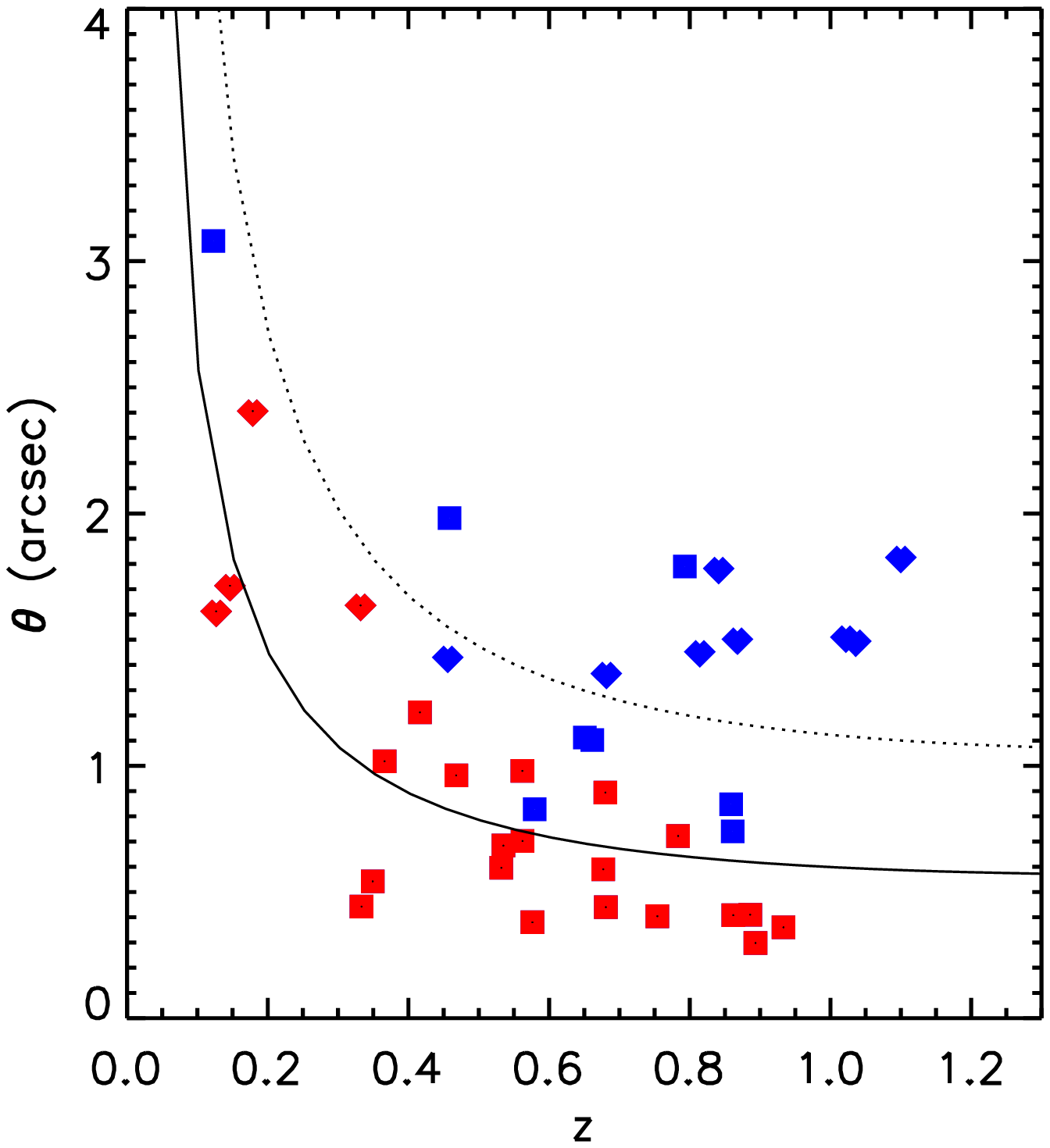}
\includegraphics[width=90mm,angle=-0]{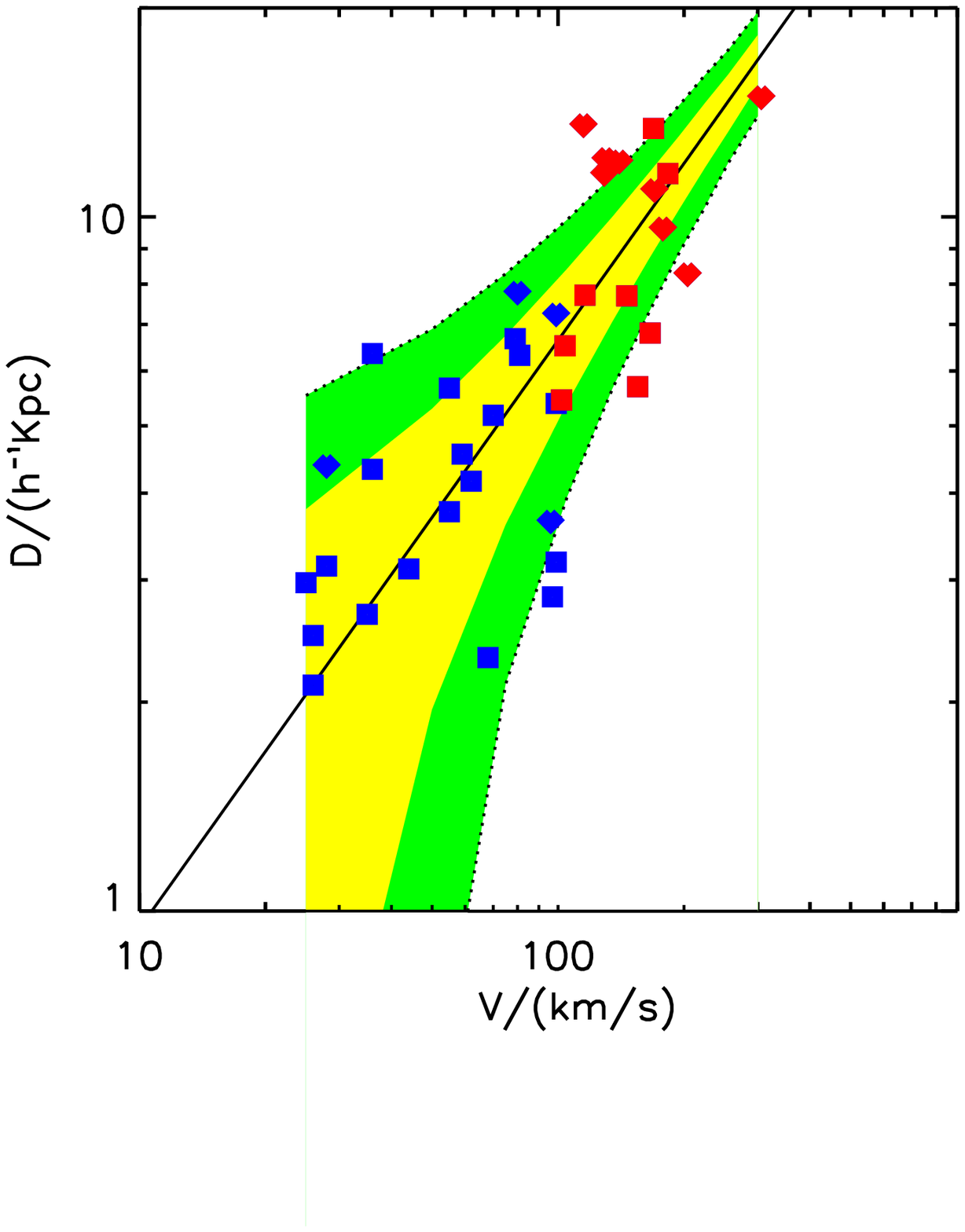}  
\caption{ {\it  Left:}
angular  diameter-redshift diagram  for  galaxies with  $0<v<100$\kms \  (red
points) and
 $100<v<200$\kms \   (blue  points). Galaxies  for    which  HST images  are
 available are indicated with
a  square. Diamonds    represent  galaxies   with  ground   photometry
\citep[EIS catalog;][]{arn01}.  The angular diameter scaling predicted
in        a             flat,         $\Lambda$-dominated    cosmology
($\Omega_m=0.3$,$\Omega_{\Lambda}=0.7$)    is  shown. The  theoretical
expectation has been derived   assuming as standard rods the   locally
($z=0$)  calibrated  half-light diameters   of   galaxies (see   Table
\ref{table2}) at the characteristic velocity corresponding to the mean
of the observed velocity distribution of galaxies  in the two velocity
ranges  considered.  These  values   are $D(z=0,v=60$ km  s$^{-1})=4.3
h_{70}^{-1}$kpc for the $S_{60}$ sample (solid line) and $D(z=0,v=143$
km s$^{-1})=9 h_{70}^{-1}$ kpc for the $S_{143}$ sample (dotted line).
{\it Right:} The inferred half-light diameter D for the $S_{60}$ (red)
and $S_{143}$ (blue) samples  of galaxies is plotted  as a function of
the rotational  velocity of galaxies.  The same cosmology as before is
assumed   for converting angles  into    linear diameters. The   local
calibration for the diameter-velocity relationship (see Eq. 3 of paper
II) is overplotted together with $1\sigma$ and $2\sigma$ uncertainties
in   the     zero-point   calibration  (shaded   area).}  \label{fig2}
\end{figure*}

After   checking the consistency   of the strategy  to select rotators
based  on the  use  of   different  spectral emission lines and different 
velocity indicators at  different redshifts, 
we now show that, by  selecting low/high velocity rotators,
we  effectively identify  distinct classes of  small/big disc galaxies
which can be used for cosmological studies.

In the  right panel of Figure  \ref{fig2} we plot the intrinsic linear
diameter  of the high redshift rotators  recovered by assuming a flat,
lambda-dominated  cosmology  ($\Omega_m=0.3$, $\Omega_{\Lambda}=0.7$,
$h_{70}=1$). The relative scatter at a given velocity is comparable to
what is found locally. In particular we observe a tighter relationship
for big rotators  and a  looser  one for smaller discs.  We are 
however comparing samples of systematically different richness. As
a  matter of  fact, because of  the  specific form of the  galaxy mass
function, the number  density of rotators  decreases as  a function of
velocity (e.g. White \& Frenk 1991, Marinoni \& Hudson 2002). This plot confirms   
that  a tight selection in  rotational velocity space   translates  
into a  tight  selection in diameters, even at high redshift.

In Figure  \ref{fig2} (left panel),  we also show the angular diameter-redshift
diagram for our sample of high redshift objects. 
While no obvious relation seems to exist between the apparent angular dimension 
and its redshift, by  separating the
sample into rotational velocity  classes ($S_{60}$ and $S_{143}$) evidence 
for this relation starts to appear; the  angle subtended by galaxies in   the low-velocity sample are
systematically lower, at any redshift, with respect to those of faster
rotators.  The  tightness of the relation becomes even
clearer when the  theoretically expected $\theta  \; vs.\; z$  scaling
relations are overplotted (the theoretical $\theta(z)$ relation assumes
the intrinsic size of the galaxies given in  Table \ref{table2} and a
flat, $\Lambda$-dominated cosmology).

\begin{figure*}          
\includegraphics[width=84mm,angle=0]{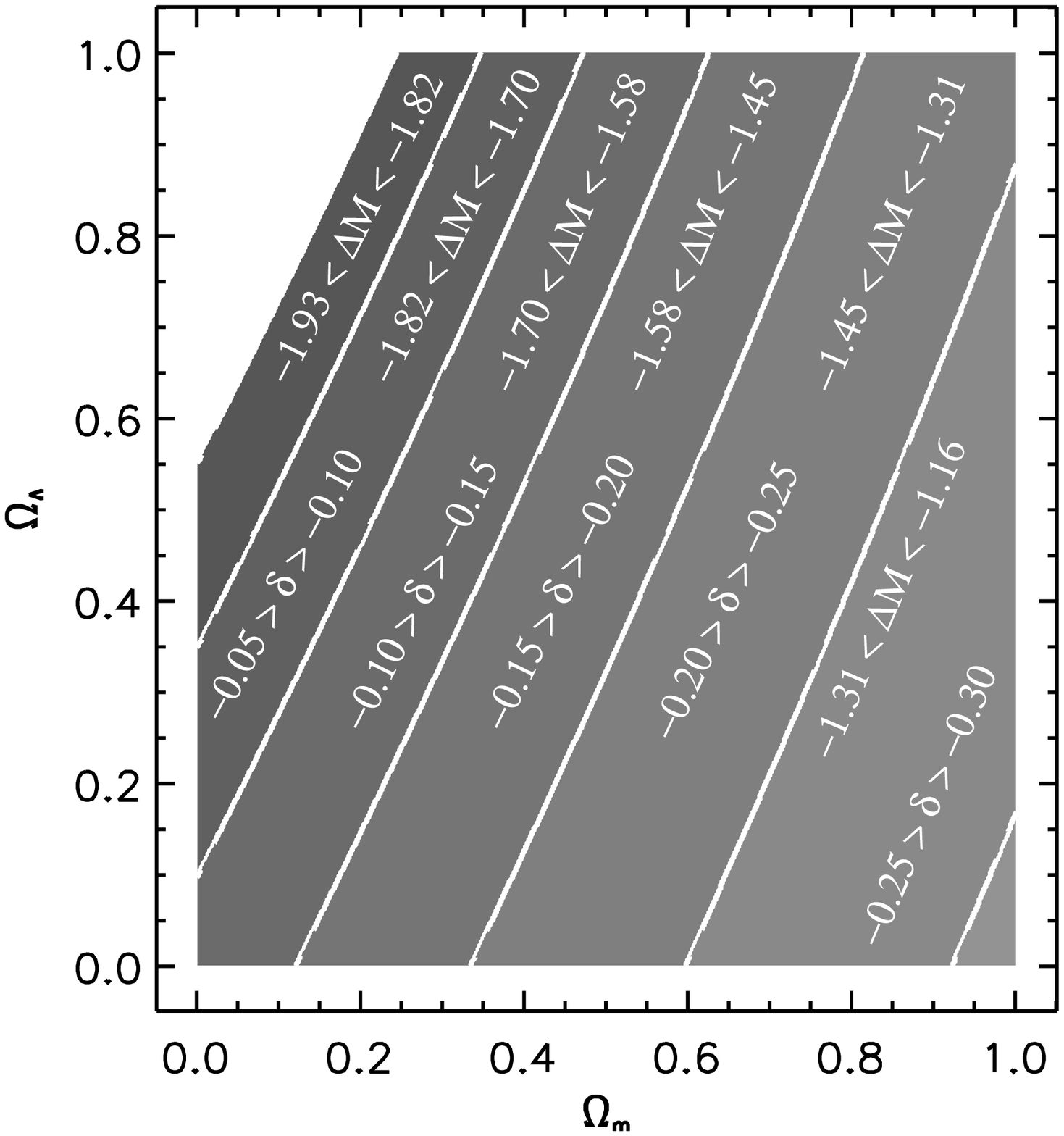}
\includegraphics[width=84mm,angle=0]{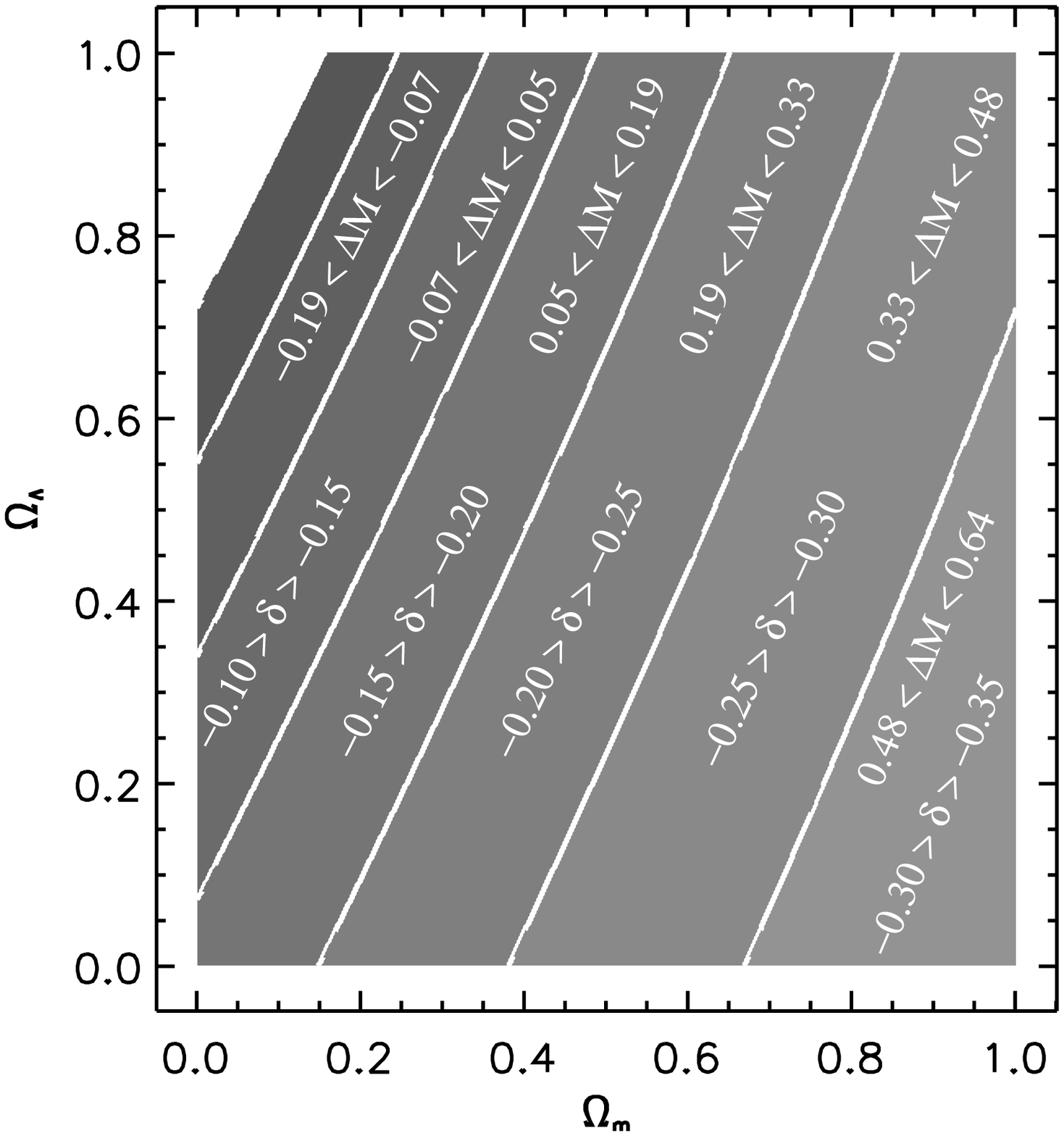}    
\caption{{\it  Left:}
Cosmology-evolution diagram for  the $S_{60}$  sample at  $z=1$.   The
cosmological plane  is partitioned with different  boundaries obtained
by solving  equation  \ref{cond5} for  different  values  of $\delta$,
i.e. of  the  relative evolution in  disc  size  at $z=1$.  Boundaries
corresponding to different relative disc evolutions from $\delta=0$ 
to $\delta=-30\%$ in steps  of $5\%$ intervals are  shown. We also show
the set of possible absolute luminosity  evolutions at $z=1$ which are
compatible with a given  set of cosmological models. These upper/lower
limits  in luminosity   evolution    have  been   derived   by   using
Eq. \ref{eqe}. Boundaries in disc relative  evolution ($\delta$) 
are uncertain  by a 23\% factor, while  luminosity evolution  boundaries are uncertain  by
$0.27$mag. {\it Right:} the same but for the $S_{143}$ sample of higher
mass objects.}  Disc relative evolution boundaries  are uncertain by a
20\% factor, while  luminosity evolution  boundaries are uncertain  by
$0.2$mag. 
\label{fig7} 
\end{figure*}

\begin{figure*}          
\includegraphics[width=90mm,angle=0]{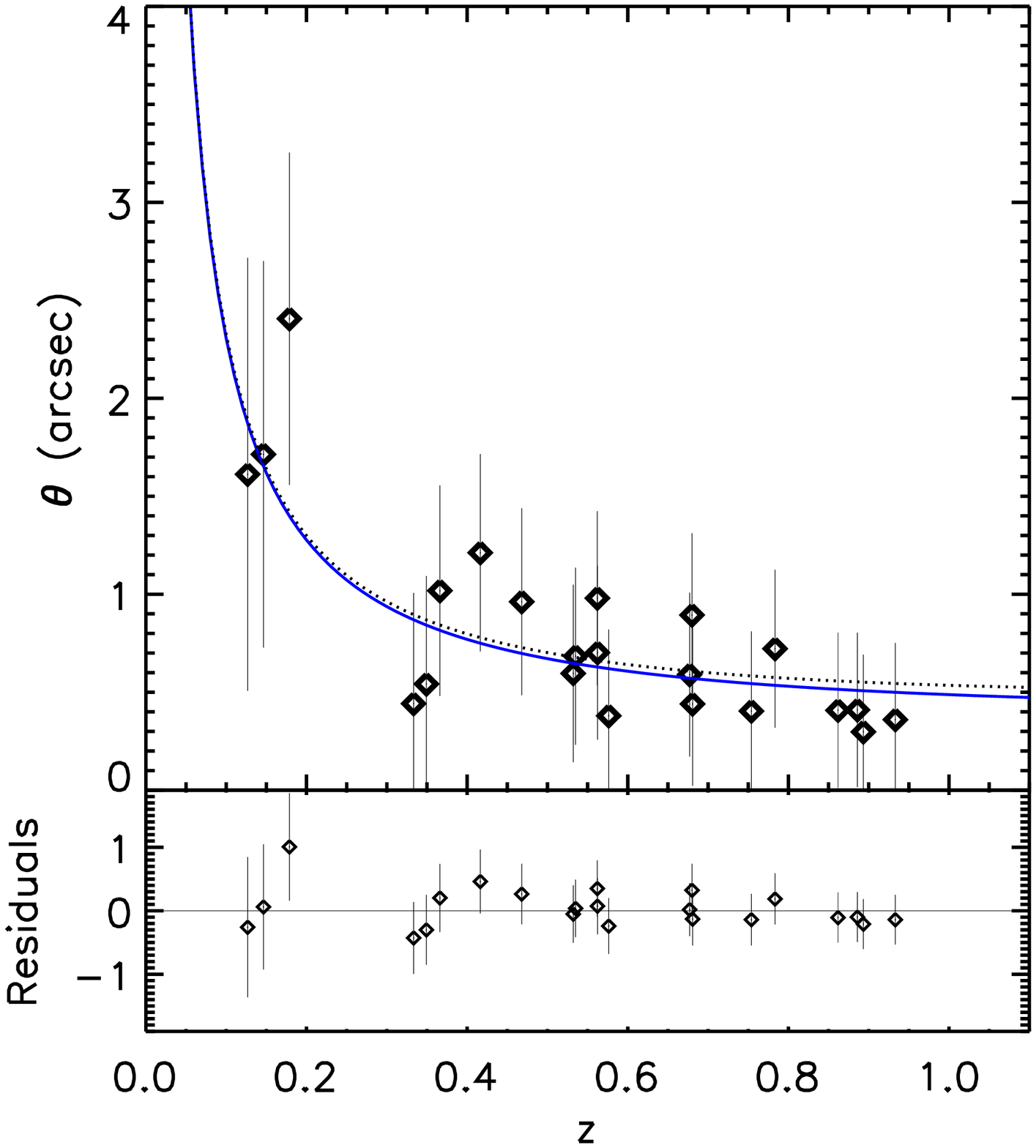}
\includegraphics[width=90mm,angle=0]{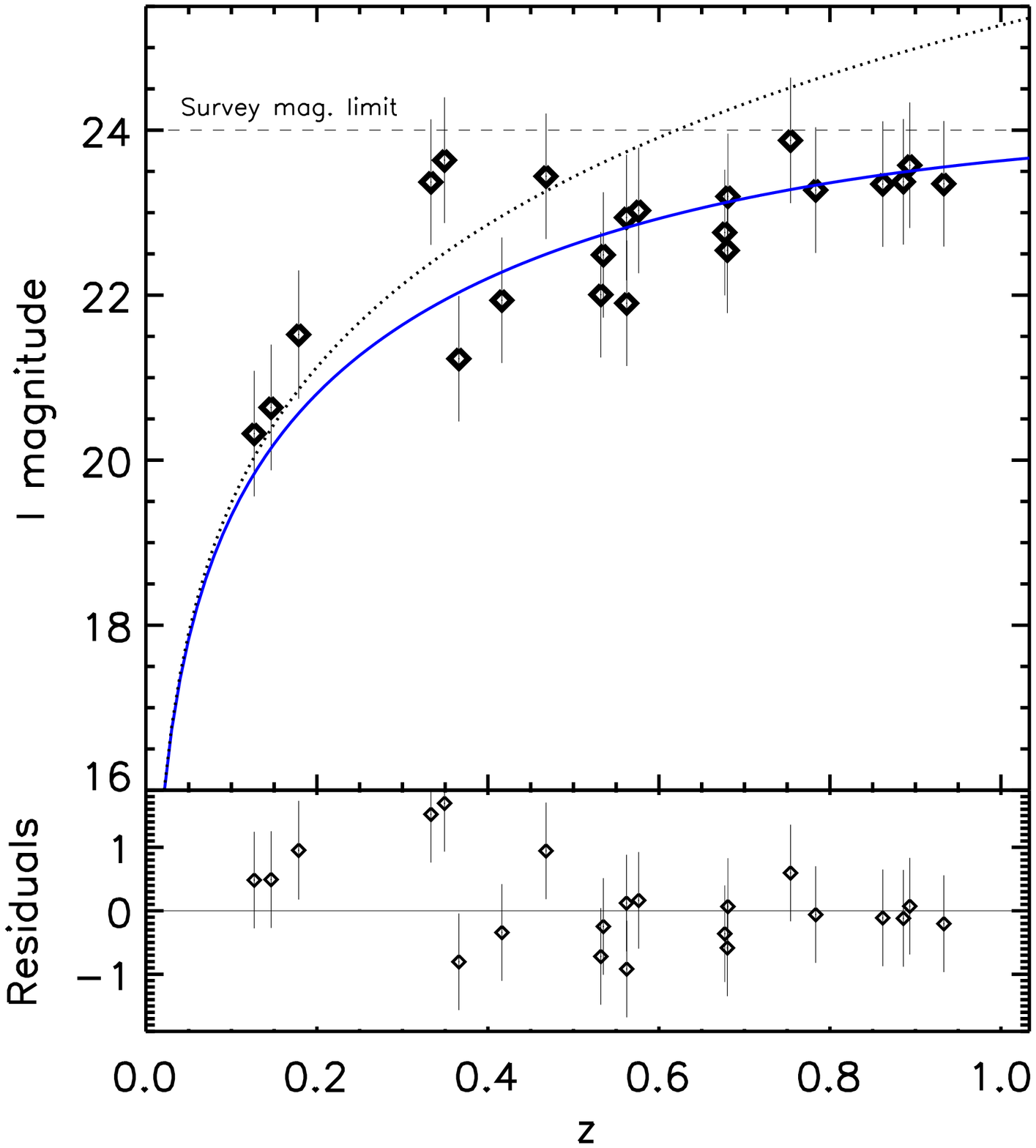}  
\caption{{\it  Left  :}
Angular diameters versus redshift for the  $S_{60}$ sample. The dotted
line represents the theoretical scaling (Eq. \ref{eqe})
 predicted assuming a standard rod
of size $D_{60}(z=0)=4.3 h_{70}^{-1}$kpc and a $\Lambda$CDM background
cosmological      model       with     parameters      ($\Omega_m=0.3,
\Omega_{\Lambda}=0.7$).  The  solid line  represent the  best  fitting
linear evolutionary model obtained  by assuming $\epsilon_D=\alpha  z$
in  Eq.  \ref{eqd}.  Errorbars  represent   the  uncertainties in  the
calibration of the local standard rod  (see Table 2) $\sigma_{D}/D\sim
0.6$.   {\it Right  :} Hubble diagram   for the $S_{60}$ rotators. The
dotted line represent  the  theoretical scaling predicted  assuming  a
standard  candle of absolute   luminosity  $M_{60}(z=0)=-18.80 +5  \log
h_{70}$ and the same  cosmology as before.   The solid line represents
the best fitting linear  evolutionary model for  luminosities obtained
by using $\epsilon_{M}=\beta z$ in Eq. \ref{eqa}.  Errorbars represent
the uncertainties in the calibration of the local standard candle (see
Table 2) $\sigma_{m}=0.75$.  } 
\label{fig3} 
\end{figure*}

\begin{figure*}          \includegraphics[width=90mm,angle=0]{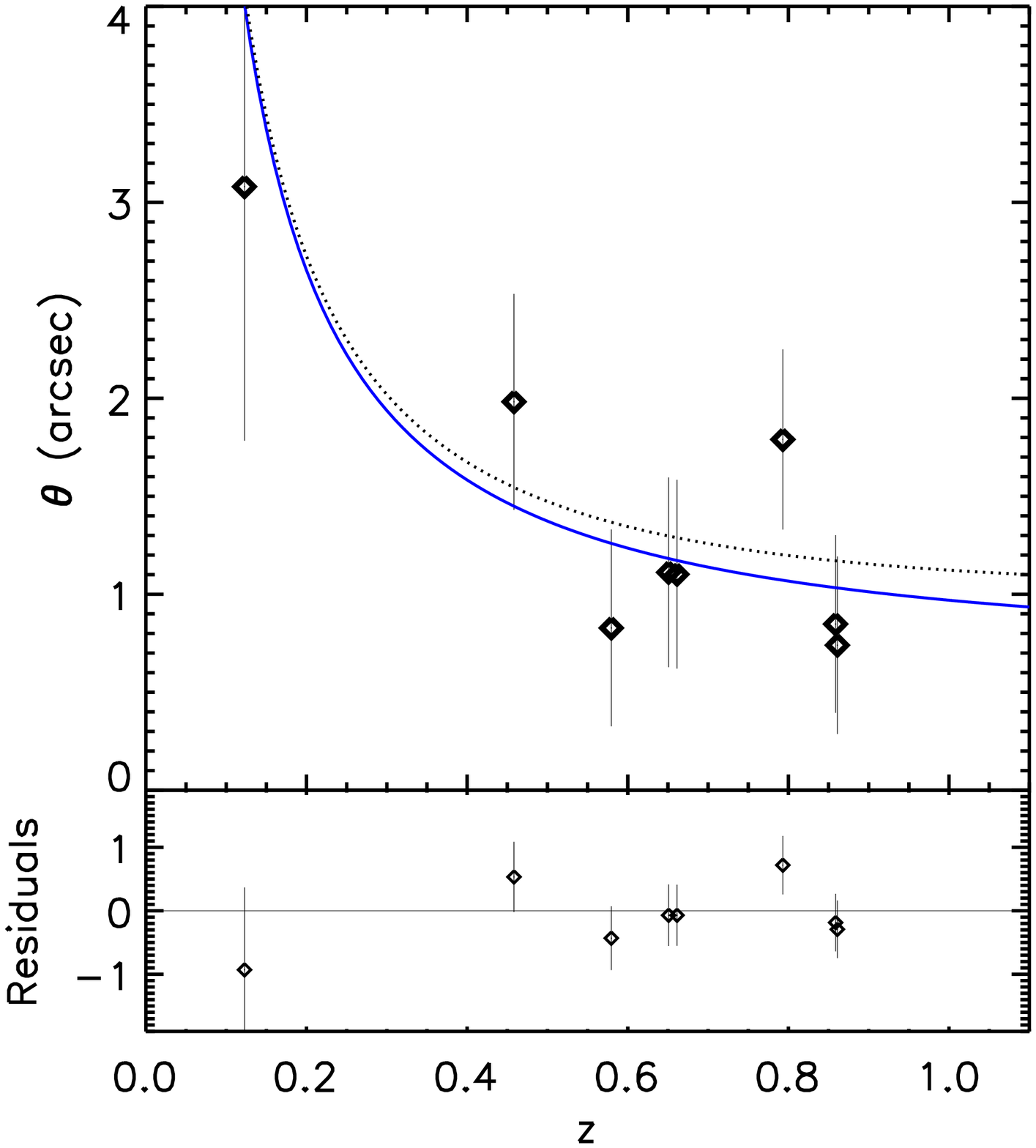}
\includegraphics[width=90mm,angle=0]{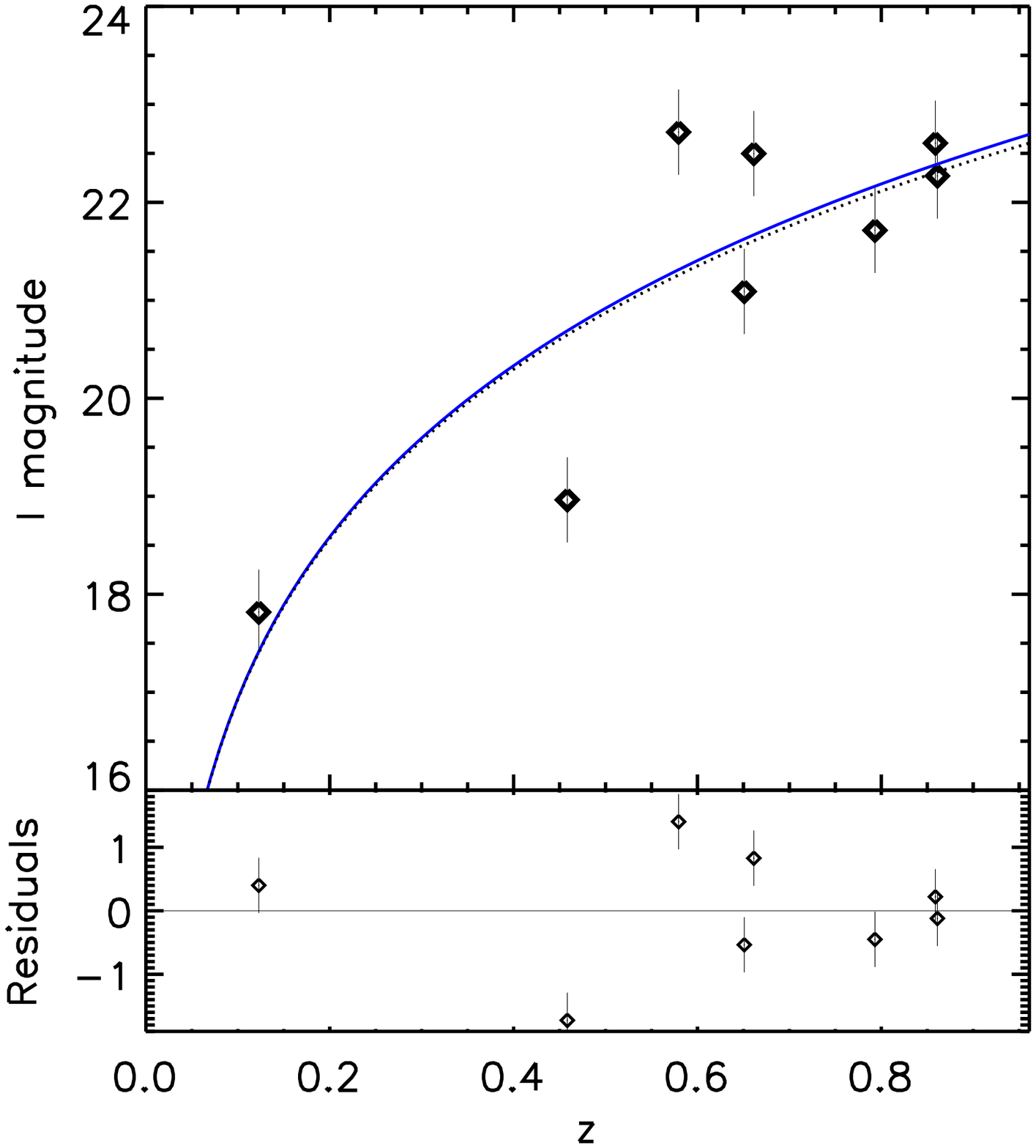} \caption{The same as  in
Figure \ref{fig3}   but   for  the  $S_{143}$ sample    of  rotators.}
Errorbars in the   angular diameter-redshift diagram ({\it  left})  represent the
uncertainties in the calibration of  the local standard rod (see Table
2)  $\sigma_{D}/D\sim  0.3$.   Errorbars  in  the Hubble-diagram ({\it
right}) represent  the uncertainties in the  calibration  of the local
standard  candle (see  Table   2)  $\sigma_{m}=   0.2$.   \label{fig4}
\end{figure*}

\section{Cosmology-Evolution diagram at $z=1$}

As shown in Paper I, if we assume that the  evolution of  discs is
linear with redshift (or can be  linearly approximated in the redshift
range of interest) and  mild (less then 30\%  at $z=1.5$), then one can
use the angular diameter-redshift test  to detect in a direct way
the eventual  presence of a dark  energy component. Since our data are
still  too  sparse  for placing any   meaningful  constraint onto this
cosmological parameter, we here  use our  sample  to  construct  the
cosmology-evolution plane (see \S 6.2 of Paper I).

This diagram allows us to visualize the set of cosmological parameters
which are  compatible  with   a  given interval   of   disc/luminosity
evolution, and {\it vice versa} how much evolution is expected given a
specific cosmology. It establishes a one-to-one correspondence between
cosmology and the amount of evolution in disc or luminosity at a given
redshift.

Given the local calibration  for diameters and magnitudes of  galaxies
within   a    particular  velocity    interval,  we     construct  the
angular diameter-redshift  and   Hubble diagrams in  any  possible cosmological
model    $\vec{p}$   spanning  the   range   $p_1=\Omega_m=(0,1)$  and
$p_2=\Omega_{\Lambda}=(0,1)$. We  then  solve for  the set  of  points
$\vec{p}$ of the parameter space which satisfy the condition
\begin{equation}    \delta_l(\bar{z})  <   \frac{\epsilon_{D}(\bar{z},
\vec{p})}{D(0)} < \delta_u(\bar{z}) 
\label{cond5} 
\end{equation} 
where $\delta_l$ and  $\delta_u$ are the lower  and upper limits in relative
disc evolution at redshift $z=\bar{z}$.
For consistency, we thus require that the amount of evolution having
to be introduced in order for both sizes 
($\epsilon_D(z,\vec{p})=D_v(z,\vec{p})-D_v(0)$)    and    luminosities
($\epsilon_M(z,\vec{p})=M_v(z,\vec{p})-M_v(0)$) to   fit
observations, be compatible with Eq. \ref{eqe} which describes the {\it
observed} evolution of  the  intrinsic mean surface brightness  within
the HLD of the objects, a cosmology-independent observable.

Solving for eqs.  \ref{eqe} and  \ref{cond5}  we can thus construct  a
self-consistent cosmology-evolution  plane,  where to any  given 
upper/lower limit for the evolution of diameters or luminosity at
$z=\bar{z}$ corresponds in a unique way a specific region  of  the
cosmological  parameter space.    In Figure  \ref{fig7}   we show  the
cosmology-evolution  diagram  for   both the  $S_{60}$  and  $S_{143}$
samples at redshift $z=1$. This plot establishes a direct link between
global properties of  the cosmological background, such as curvature,
dark  matter  and  dark  energy  content, and  the  local  structural
parameters of rotators.

Let's assume that the luminosity  of $v=200$ \kms \ rotators
cannot be fainter at $z=1$, which means that the light
output of  high redshift rotators hosted in  dark matter halos  of
$v=200$  \kms \  cannot be smaller than that emitted  by present day
galaxies hosted in such halos. This can be expressed as the following 
boundary condition for the luminosity evolution of the fast rotators:
$\Delta M(z=1)   \leq 0$.  Therefore, we assume that the luminosity 
produced per unit mass is declining (or at most constant) since $z=1$, and 
since we  are considering  halos of  similar mass, that galaxies as a whole have been 
fading away. By
inspecting the cosmology-evolution diagram for the $S_{143}$ sample we
can conclude, using this {\it a-priori} constraint, that a flat, matter-dominated 
 cosmology ($\Omega_m=1$) is excluded at a
confidence level  of $\sim  3  \sigma$.  Even more  interestingly, the
$\Delta M(z) \leq  0$ constraint allows us  to conclude  that an open
cosmology with low mass density ($\Omega_m \sim 0.3$) and with no dark
energy contribution ($\Omega_{\Lambda}$)  is excluded
at a confidence level greater than $1 \sigma$.

We stress that these cosmological conclusions are drawn by assuming
that, whatever the strength of the luminosity evolution of galaxies with redshift, 
this evolution cannot lead to the brightening $v=200$ \kms  \  rotators from  $z=1$ to the
present    time.   We  include  evolution  in   our  analysis from the
beginning, and we  only reject {\it a-posteriori}  cosmological models
that are associated at a  particular cosmic epoch  ($z=1$ in our case)
with unlikely galaxy evolutionary models.

\begin{figure}          \includegraphics[width=90mm,angle=0]{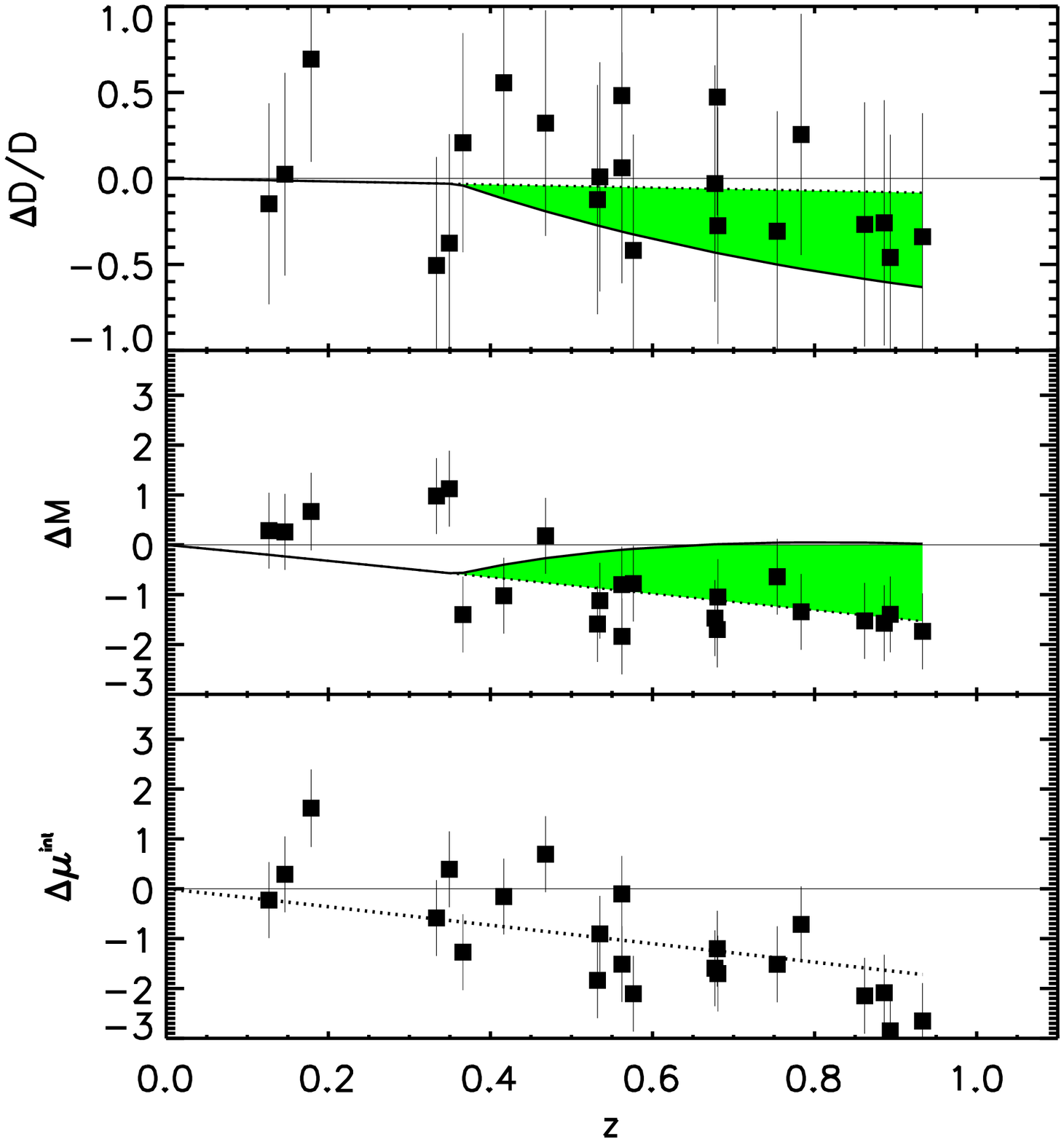}
\caption{{\it  Top:} relative evolution in  the diameter  size for the
$S_{60}$ sample  of  low velocity  rotators.  We assume the  zeropoint
diameter  normalization   quoted  in  Table   \ref{table2}   and  a  a
$\Lambda$CDM cosmological  framework.  Errorbars represent   1$\sigma$
scatter in the calibration of the  local standard rod. The dotted line
represent  the best fitting linear   evolutionary model for  diameters
obtained by assuming $\epsilon_D=\alpha z$ in Eq. \ref{eqd}. The solid
line represents the upper limit  on relative disc evolution  estimated
by  correcting for  the  Malmquist  bias  affecting  the   data.  {\it
Center:}  Evolution  of  the absolute magnitude   of   galaxies in the
$S_{60}$ sample and   in a  $\Lambda$CDM cosmological   framework.  We
assume as standard luminosity the mean absolute  magnitude of a sample
of similar rotators at $z \sim 0$ (see Table \ref{table2}).  Errorbars
represent  1$\sigma$ scatter in the calibration  of the local standard
candle. The dotted line represent the best fitting linear evolutionary
model for  luminosities  obtained by  using $\epsilon_{M}=\beta  z$ in
Eq.  \ref{eqa}.    The solid  line represents   the   inferior limit on
absolute magnitude evolution estimated by correcting for the Malmquist
bias affecting the  data.   {\it Bottom:}  evolution in  the intrinsic
surface  brightness within the half-light  diameter. This evolution is
independent of  the particular  cosmological background.    The dotted
line represents   the  combination (Eq. \ref{eqe}) of the best
fitting diameter  and luminosity evolution  functions.}   \label{fig5}
\end{figure}

\begin{figure}          
\includegraphics[width=90mm,angle=0]{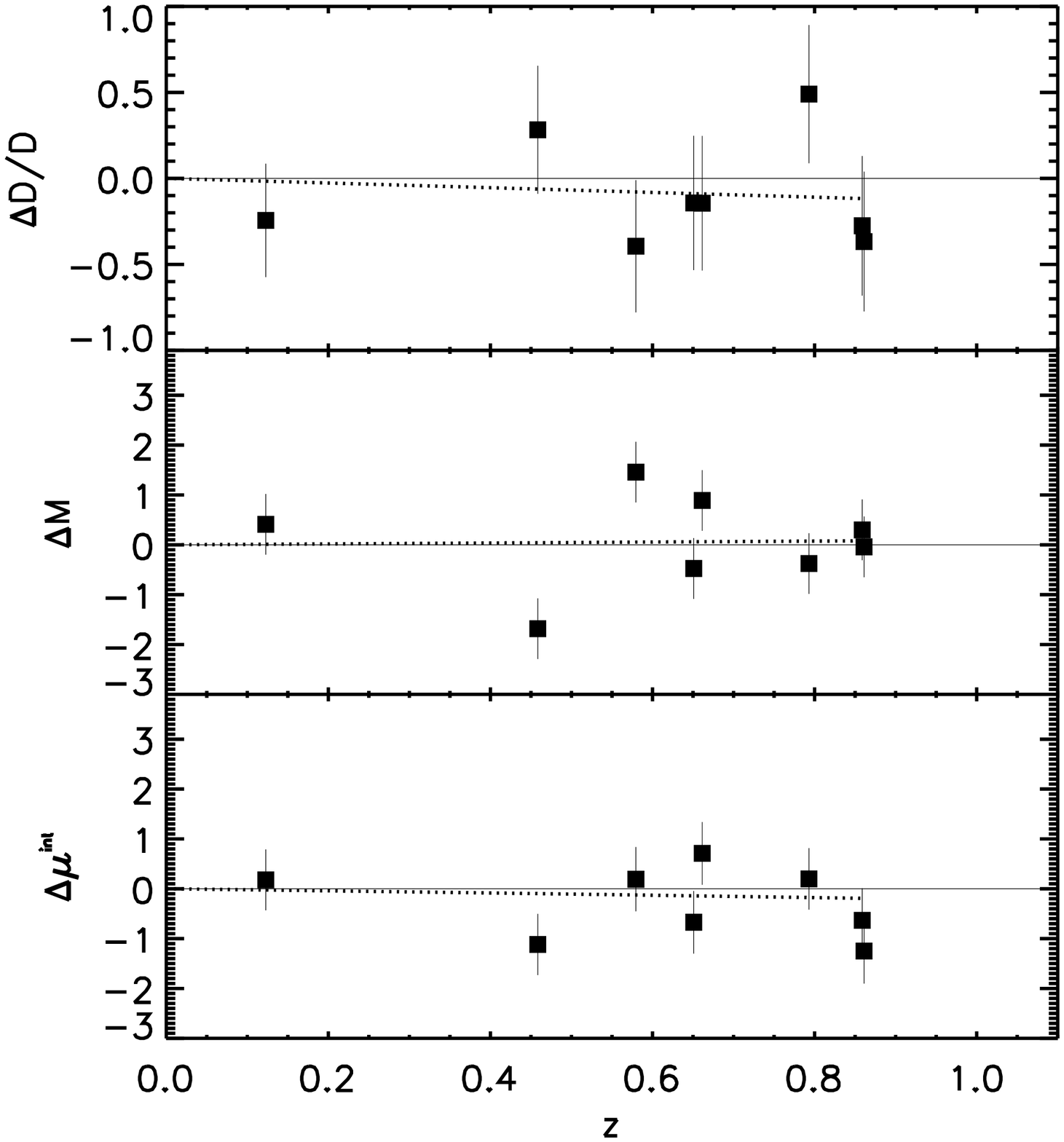}
\caption{Same  as Figure  \ref{fig5}  but for  the $S_{143}$  sample.}
\label{fig6} 
\end{figure}

\section{Diameter    and Luminosity  evolution    in  a $\Lambda  CDM$
cosmological model}

Insights into the  mechanisms  of galaxy evolution   are traditionally
accessible through the study of disc galaxy scaling relations, such as
the    investigation  of      the   time-dependent  change     in  the
magnitude-velocity          (Tully-Fisher)      relation        (e.g.,
\citet{vog96,bohm04,bam06}), of  the  magnitude-size  relations (e.g.,
\citet{lil98,sim99,bou02,bar05}),     or    of  the  disc  ``thickness"
\citep{res03,elm05}.     Yet owing to   sample  selection effects, and
differences in analysis techniques, these studies  have come to widely
divergent  conclusions. In this study,  we have explored and adopted a
different approach: we  infer information  about  size and  luminosity
evolution  of   galaxies  by  constructing their   respective  angular
diameter-redshift and Hubble diagrams in a fixed reference cosmology.

In  Figure \ref{fig3} and \ref{fig4}  we show the angular diameter-redshift
and Hubble diagrams for  the $S_{60}$ and $S_{143}$ samples, respectively.
The expected  scaling in the  flat, $\Lambda$-dominated cosmology with
$\Omega_m=0.3$ and $\Omega_{\Lambda}=0.7$  is shown  together with the
best fitting function obtained by   assuming a simple linear  redshift
evolution for both  diameters and absolute magnitudes.  Both diameters
and angles for the 2 velocity samples are normalized at $z=0$ by using
the values derived in Paper II and shown in Table \ref{table2}.

The disc and  luminosity evolution  in a  $\Lambda CDM$ cosmology  for
both small  and  large rotators  is   shown in  Figure \ref{fig5}  and
\ref{fig6}, respectively.  In these figures, we also show how these
theoretically-derived evolutionary patterns  combine together  to give
the evolution of the intrinsic surface brightness  (see Eq. 5), and how
this last quantity compares to the observed one, which as stated earlier does 
not depend on the adopted cosmological model.

As stressed in Paper I, the test  should be performed with big rotators
i.e.  using  bright  candles   whose  selection   is  unbiased  in   a
flux-limited spectroscopic survey. Our  sample of $v=200$ \kms \ meets
this criteria.  However, because of  the specific form of  the  galaxy
luminosity function, our preliminary   sample is dominated by
small rotators, whose magnitude distribution could be affected by the
Malmquist bias:  the observed distribution  of galaxies might not
include the  fainter  tail   of  members having  a  rotation  velocity
satisfying our selection  criteria ($0<v<100$). Even though he GOODs catalog is
virtually unbiased in surface brightness selection for the magnitude 
range considered in this paper,  our measurements of surface 
brightness evolution could be biased simply because of the flux cut at I=24 
characterizing our sample. 
One could in principle  miss low surface brightness galaxies of the same
size as those observed, simply because their magnitudes are fainter
then the survey limit. However, using the low redshift SFI++ sample we have checked that 
galaxies that are on the faint end tail of the magnitude distribution 
also tend to be the smaller discs. Therefore, no Malmquist effect is 
expected  to contaminate the observed intrinsic surface brightness evolution.

The spectroscopic  survey in  the CDFS region is flux-limited    at
$I<24$. Since  the  standard candle  of  the  $S_{60}$ sample  has  an
absolute luminosity  $M_{60}=-18.8 + 5\log  h_{70}$ while the brighter
luminosity sampled at $z=1$ is $M=-20.2+5\log  h_{70}$,
we could be overestimating the observed
evolution  in   luminosity. The  fact   that   we see   in   our  I=24
magnitude-limited sample rotators with $v=100$ \kms \  at $z=1$ can be
interpreted in  two  different ways:  i)   these rotators were
effectively much brighter  in  the past,  or ii)  we  only sample  the
brightest objects, scattered around  the standard absolute value.  To
address the latter, we correct our  results for any possible Malmquist
bias.

Let's  consider the difference  between the survey  flux limit and the
theoretically-predicted best fitting function  to  the  observed
magnitude distribution,
$\Delta=24-m^{bfit}$. A simple estimate  of the Malmquist  bias
is obtained  by assuming that the best fitted
apparent magnitude  is  systematically overestimated as  a function of
distance by the additive quantity $3\sigma-\Delta(z)$, where we assume
that the  scatter in the standard  candle calibration is constant as a
function of redshift.  In other words, we  assume that the galaxies we
see are the brighter subset of standard rods whose luminosity scatters
around $M_{60}$.

We have implemented    this correction consistently   both for  galaxy
luminosities and diameters. The incidence of the Malmquist bias on our
conclusions is graphically  shown in  Figure  6.   Due to the   strong
influence of  the Malmquist  correction term,  the  observed disc  and
luminosity evolution   for   the  slow rotators  sample   $S_{60}$  is
compatible with the  following diametrically opposite interpretations:
i) data are affected  by the Malmquist bias  and therefore small discs
have undergone almost no   luminosity   evolution but a strong    size
evolution  (they were nearly 50  \%  smaller at  $z=1$ than at present
epoch), or ii) data are unaffected by the Malmquist bias and the small
discs  have undergone strong  luminosity  evolution but not much  size
evolution since $z\sim 1$.

Since there is marginal evidence that  the scatter around the expected
disc and luminosity evolution is decreasing as redshift increases (see
Figure  6),  we take  a  conservative  position  and  assume,  in the
following,  that our $S_{60}$ sample  is affected by the Malmquist bias.
Only a sample of small rotators selected in a magnitude-limited survey
deeper than the CDFS will allow  to unambiguously resolve the issue by
differentiating between the two opposite scenarios. However, we  stress
again  that the large rotators  sample $S_{143}$ does  not suffer from
Malmquist bias selection effects.

\section{Discussion on the Evolution of Structural Parameters}

Assuming  a $\Lambda CDM$ cosmology, several conclusions can be
reached   about the    evolution   of the     structural parameters of
fixed-velocity rotators.  \begin{enumerate}
\item The  surface  brightness evolution   of  discs is  significantly
      different
for the two populations, $S_{60}$ and $S_{143}$: $\Delta \mu=-1.90 \pm
0.35$ mag/arcsec$^2$ for  the  slow  rotators ($S_{60}$) and   $\Delta
\mu=-0.25 \pm 0.27$ mag/arcsec$^2$ for the fast rotators ($S_{143}$).
\item  The fast rotators  show neither significant size nor luminosity
      evolution since $z=1$.
\item Under  the conservative assumption that   most of the luminosity
      difference over redshift
for the small  rotators is due to  the Malmquist bias, they  appear to
have gone through   a  significant size  evolution  and  no luminosity
evolution since $z=1$.  
(however the opposite is true in the limiting case in which Malmquist bias 
minimally affects our low-velocity data. In this case,  small
discs  have undergone strong  luminosity  evolution but not much  size
evolution since $z\sim 1$.)

\end{enumerate}

The results presented in  the previous section highlight the potential
of  the geometrical  tests,     not only to   constrain   cosmological
parameters, but  also  to derive information  about  the  evolution of
galaxy structural parameters. However, the extent of the analysis that
can be performed  at this point is significantly  reduced by the small
sample of galaxies available with deep  photometry and high resolution
spectroscopic measurements,  and  by  the  limiting magnitude   of the
CDFS. This  magnitude limit    introduces a potentially    very strong
Malmquist bias for the sample of small discs, $S_{60}$, which prevents us
from determining  if the   surface brightness increase   of  $\sim1.9$
mag/arcsec$^2$   of  these  galaxies   at $z=1$ is  due   to  a strong
luminosity  or size  evolution   (or  a  combination  of  both).  This
distinction  has not been  made  by most  previous studies either, but
could be achieved in the future using the strategy proposed here and a
deeper, more complete galaxy sample.

Because at  the  limiting magnitude of I=24  of  the VIMOS spectroscopic
survey the GOODs photometric catalog is unbiased in surface brightness
selection,   conclusions  can  however be  reached  about  the surface
brightness evolution of discs.  While we find strong evolution for the
small rotators, the  large rotators seem  to have retained  a constant
surface  brightness  since $z=1$.   The  evolution  of more  than  one
mag/arcsec$^2$ at $z=1$ for the small discs is consistent with results
of     previous     studies    of    the     magnitude-size   relation
(e.g.    \citet{sch96,for96,lil98,roc98,sai05,bar05}), or   of     the
magnitude-velocity (Tully-Fisher) relation
(e.g. \citet{mil03,bohm04,bam06}).   While  others  report very   little  or  no surface
brightness     evolution  (e.g. \citet{sim99,rav04}),    \citet{bar05}
reconciles this discrepancy by considering the different data analysis
techniques.  It  also seems  likely  that some  of these  results  are
affected by the selection criteria applied.  For example, some authors
selected their  samples  on the  basis of   blue colors \citep{rix97},
strong emission line   equivalent widths \citep{sim98}, or large  disc
sizes \citep{vog96}. The two former criteria prefer late-type spirals,
whereas the latter criterion leads to the overrepresentation of large,
early-type  spirals. With our   strategy,   based on a   spectroscopic
follow-up  of objects with  OII emission-lines  selected from a purely
flux-limited redshift survey, biases are largely reduced.

In  \S3.1, we have  shown that  the  two classes  of galaxies at  high
redshift with rotation velocity estimated  on the  basis of their OII linewidths 
represent  in  an  unbiased way the  progenitors  of local  discs 
whose velocity is inferred using the $H{\alpha}$ rotation curves.
We  stress that this statement  does not imply  that
every high redshift galaxy with the same rotational velocity 
as a local galaxy  is its  direct progenitor. 
Due to  the prevalence   of  mergers,
interactions and accretion phenomena in the  past, this is actually an
unlikely scenario.  What the $\eta$-test guarantees is  that
the high- and low-$z$ samples  represent the
same populations of rotators with nearly the same mass.   
While interactions were more frequent in the past, are known to lead to 
the onset of star
formation events and could  therefore  provide an explanation  of  the
excess in luminosity   of the small discs   at high $z$, all the   ACS
images were examined and show that all the  galaxies in the sample are
not undergoing merger events.

Under the  hierarchical   scenario for  the growth  of  structure, the
following scaling relation for  the  disc scalelength, $R_d$,  of dark
matter systems is predicted \citep{mo98}: \[ R_d \approx
8.8   h^{-1}  {\rm kpc}   \left(  \frac{\lambda}{0.05} \right)  \left(
\frac{V_c}{250  {\rm   km   s}^{-1}} \right)  \left(  \frac{H(z)}{H_0}
\right)^{-1} \left(   \frac{j_d}{m_d}  \right) \]    where
$\lambda$ is  the  disc spin  parameter, $V_c$  the circular velocity,
$j_d$ the angular  momentum of the  disc as a  fraction of that of the
halo, and   $m_d$ is the  disc   mass as a  fraction  of  that  of the
halo. Since there is no dependency  of $\lambda$ on redshift and since
$j_d/m_d=1$  if  discs are  formed while  conserving  specific angular
momentum, the  disc scalelength of dark  matter systems having a given
circular velocity  scales as  \begin{equation} R_d  \propto H^{-1}(z),
\label{rd}  \end{equation}  where $z$  corresponds to  the redshift of
formation of the galactic discs. Under the fairly safe assumption that
the   diameter of  the stellar disc    of  a galaxy    scales with the
scalelength    of  its  dark   matter   halo,   the  diameter  of  our
velocity-selected   standard rods  should   also evolve  as $D \propto
H^{-1}(z)$. This relation therefore predicts that discs forming at the
present epoch  should  be larger by a  factor  of 1.8 than  those that
formed  at  $z=1$.  This estimate  agrees   with our  most pessimistic
scenario for the Malmquist bias, which asks for  the small discs to be
larger by a factor of two at $z=0$.

Interestingly,  the fact that no size  evolution is seen for the large
rotators  may tell us  something about the time  of formation of these
systems. In Eq. \ref{rd}, the dependency of the radius $R_d$ is on the
value of the Hubble constant {\it at the epoch of last assembly of the
discs}. This corresponds to the last time when there was a significant
reshuffling of   the disc  (i.e.  the  last    incidence  of  a  major
merger).  Since the discs in our   $S_{60}$ sample show size evolution
that  is consistent with   Eq.  \ref{rd}, the  redshifts  at which  we
observe them probably coincide with  the epochs at which these systems
are still forming. Due to the paucity of the galaxy sample at hand, no
firm conclusion can be reached at this point,  but there are some 
hints that the large  discs were already in
place   by $z=1$ while  most of  the  small discs have assembled since
then. There is tantalizing evidence of a similar effect in the scaling
relation between disc  scalelength and rotation velocity observed  for
nearby galaxies, where the smallest discs appear to be consistent with
an epoch of last  formation at $z<0.5$ while  this  is pushed  back at
$0.5<z<1.0$ for the larger systems \citep{spe05}.

Note that a relation similar to Eq.  \ref{rd} also exists for the disc
surface density, $\Sigma_0$. However,  it is not as straightforward to
extend  that relation to predict   the behavior of surface brightness,
because   of     its  dependency     on  the    stellar  mass-to-light
ratio. Therefore, a similar interpretation   can not be made for   the
observed surface  brightness evolution of   the  discs, as it  is  not
possible  to   disentangle the  combined     effects of redshift   and
mass-to-light ratio evolution.

Finally we note that fast rotators cannot have stopped forming stars at 
least over the epochs explored in this study.
The constancy of their luminosity up to  $z=1$ can be explained in
terms of star-formation   activity continuously on-going for a  Hubble
time.

\section{Conclusions}

The  goal of  this pilot  observational program  
is  to investigate  the  relationship
between global properties of the  universe (geometric curvature,  dark
matter   and dark energy  content) and  local structural parameters of
disc galaxies (disc linear size and absolute luminosity).  To this purpose 
we apply the angular size-redshift  
test and the Hubble diagram at the same time to the same class of 
standard objects, namely, velocity-selected disc galaxies.
As such, it presents  one of the first
attempts  to  investigate if specific subsamples of high redshift  galaxies
 can be used as cosmological tracers complementary to SNIa and CMB observations.
This approach allows us to construct a cosmology-evolution diagram at 
redshift $z=1$, the chart that allows for the mapping of the  
cosmological parameter space onto
the disc galaxy structural parameter space (diameter, luminosity
and surface brightness).
Assuming prior knowledge about disc evolution, this diagram allows us to 
draw some interesting cosmological conclusions.
If we assume that  the    absolute magnitude evolution   is
constrained  to be   negative at  $z=1$ (i.e.  it  is impossible  that
$v=200$ \kms \  rotators were  fainter at  $z=1$ than  at the  present
time, which  means that their luminosity per unit mass was higher in the past), 
we find with the data at hand that: 
\begin{enumerate}
\item a flat matter-dominated cosmology ($\Omega_m=1$)
is excluded at a confidence level of $2 \sigma$.
\item an open  cosmology with low mass  density ($\Omega_m  \sim 0.3$)
       and no
dark energy contribution  ($\Omega_{\Lambda}$) is  excluded by present
data at a confidence level of $1 \sigma$.  \end{enumerate}

On the other hand, by fixing the background cosmological model, the cosmology-evolution
diagram allows us to investigate the evolution in the structural parameters 
of disc galaxies which are hosted in dark matter halos of {\it similar mass}.
Assuming a $\Lambda CDM$ model, we find that:
\begin{enumerate}
\item while small mass galaxies  go through a strong surface brightness
      evolution of
$-1.90 \pm 0.35$ mag/arcsec$^2$  since $z=1$, larger discs only
evolve by $-0.25 \pm 0.27$ mag/arcsec$^2$,
\item  under  the  assumption that our  sample  of small discs is
      affected by the Malmquist bias, this surface brightness
evolution is  caused by an increase  in the size of the fixed-velocity
rods by a factor of two from $z=1$ to the present epoch,
\item  discs hosted in more massive halos, which are not  affected by the Malmquist bias,
      show neither size nor luminosity evolution, suggesting
that  they finished assembling  before $z=1$, unlike the smaller discs
that are still undergoing formation at $z<0.5$.  \end{enumerate}

We  conclude that the luminosity evolution   observed is coherent with
the emerging  picture  of a differential  star  formation  history for
galaxies of different masses \citep{jun05}.    These results are  also
consistent  with  the growth of  structure  predicted  in the universe
described by the concordance cosmological model \citep{mo98,bow06}.

In this  preliminary study we  are still limited   by the small number
statistics affecting our sparse sample.  While with a larger sample of
high resolution spectra and images one can  detect in a direct way the
eventual presence of a dark energy component (see Paper I), it was not
possible to apply the angular diameter-redshift test 
and put a  constraint on its amplitude  and on its equation of
state parameter $w$ with the amount of data  currently available.  For
the same reason, galaxies were separated in only two velocity bins
to construct the cosmology-evolution diagram. This limits the class
of mass-selected objects for which we can trace evolution across 
different cosmic epochs. The availability of a larger sample
will  allow finer velocity  bins,  and therefore less  scatter  in the
results.

To conclude, we reiterate that the rotational velocity of distant galaxies,  when interpreted as a size
(luminosity) indicator,  may be used as  an interesting tool to select
high redshift standard rods (candles).   
Though the  power  of geometrical    tests to constrain    fundamental
cosmological parameters has long been recognized, only with the recent
large,  deep redshift surveys   have  their implementation  been  made
possible. With  only a limited   amount  of  data but   a novel and
physically justified technique to select  standard rods/candles, we have shown
that these tests can give useful insights 
not only  on the value of fundamental cosmological parameters, 
but also the time  evolution of fundamental galaxy observables in mass-selected disc rotators.

\section*{Acknowledgments}

We wish to thank the referee for many useful comments and suggestions.
This work has been partially  supported by NSF grants AST-0307661  and
AST-0307396 and was done while AS  was receiving a fellowship from the
{\it  Fonds  de   recherche sur   la Nature  et   les  Technologies du
Qu\'{e}bec}.  CJW is  supported by the  MAGPOP Marie Curie EU Research
and  Training  Network.    KLM  is  supported    by  the  NSF    grant
AST-0406906. The  VLT-VIMOS   observations have  been  carried out  on
guaranteed time (GTO)  allocated by the  European Southern Observatory
(ESO) to the VIRMOS consortium,  under a contractual agreement between
the Centre National de la Recherche  Scientifique of France, heading a
consortium  of  French and  Italian  institutes,  and ESO,  to design,
manufacture and test the VIMOS instrument.

\label{lastpage}

\end{document}